\documentclass[11pt]{iopart}
\pdfoutput=1
\usepackage{graphicx}% Include figure files
\usepackage{dcolumn}% Align table columns on decimal point
\usepackage{bm}% bold math\footnote{•} 
\usepackage{iopams}
\usepackage{color}

\begin{document}

\title{First international comparison of fountain primary frequency standards via a long distance optical fiber link}
\author{J~Gu\'{e}na$^{1}$, S Weyers$^{2}$, M~Abgrall$^{1}$, C Grebing$^{2}$, V~Gerginov$^{2,\star}$, 
P Rosenbusch$^{1}$, S~Bize$^{1}$, B~Lipphardt$^{2}$, H Denker$^{3}$, N Quintin$^{4}$, S M F Raupach$^{2}$, D Nicolodi$^{1,\star}$,
F Stefani$^{1,4}$, N Chiodo$^{4}$, S Koke$^{2}$, A Kuhl$^{2}$, F Wiotte$^{4}$, F Meynadier$^{1}$, E Camisard$^{5}$, C Chardonnet$^{4}$, Y~Le Coq$^{1}$, M Lours$^{1}$, G Santarelli$^{6}$, A Amy-Klein$^{4}$, R~Le Targat$^{1}$, O Lopez$^{4}$, P~E~Pottie$^{1}$, G Grosche$^{2}$}

\address{$^1$LNE-SYRTE, Observatoire de Paris, PSL Research University, CNRS, Sorbonne Universit\'{e}s, UPMC Univ. Paris 06,
 61 avenue de l'Observatoire, 75014 Paris, France\\
$^2$Physikalisch-Technische Bundesanstalt (PTB), Bundesallee 100, 38116 Braunschweig, Germany\\
$^3$Institut f\"{u}r Erdmessung, Leibniz Universit\"{a}t Hannover, Schneiderberg 50, 30167 Hannover, Germany\\
$^4$Laboratoire de Physique des Lasers, Universit\'{e} Paris 13, Sorbonne Paris Cit\'{e}, CNRS, 99 Avenue Jean-Baptiste Cl\'{e}ment, 93430 Villetaneuse, France\\
$^5$R\'{e}seau National de t\'{e}l\'{e}communications pour la Technologie, l'Enseignement et la Recherche,
23$-$25 Rue Daviel, 75013 Paris, France\\
$^6$Laboratoire Photonique, Num\'{e}rique et Nanosciences, Institut d'Optique Graduate School, CNRS, Universit\'{e} de Bordeaux, 33400 Talence, France\\
${^\star}$Present address: National Institute of Standards and Technology (NIST), 325 Broadway, Boulder, CO 80305, USA  }
%\ead{jocelyne.guena@obspm.fr}

\begin{abstract}
We report on the first comparison of distant caesium fountain primary frequency standards (PFSs) via an optical fiber link. The 1415\,km long optical link connects two PFSs at LNE-SYRTE (Laboratoire National de m\'{e}trologie et d'Essais - SYst\`{e}me de R\'{e}f\'{e}rences Temps-Espace) in Paris (France) with two at PTB (Physikalisch-Technische Bundesanstalt) in Braunschweig (Germany). For a long time, these PFSs have been major contributors to accuracy of the International Atomic Time (TAI), with stated accuracies of around $3\times 10^{-16}$.  They have also been the references for a number of absolute measurements of clock transition frequencies in various optical frequency standards in view of a future redefinition of the second. The phase coherent optical frequency transfer via a stabilized telecom fiber link enables far better resolution than any other means of frequency transfer based on satellite links. The agreement for each pair of distant fountains compared is well within the combined uncertainty of a few 10$^{-16}$ for all the comparisons, which fully supports the stated PFSs' uncertainties. The comparison also includes a rubidium fountain frequency standard participating in the steering of TAI and enables a new absolute determination of the $^{87}$Rb ground state hyperfine transition frequency with an uncertainty of $3.1\times 10^{-16}$. 

This paper is dedicated to the memory of Andr\'{e} Clairon, who passed away on the 24$^{th}$ of December 2015, for his pioneering and long-lasting efforts in atomic fountains. He also pioneered optical links from as early as 1997.       
\end{abstract}

%\footnote[$^{\star}$]{${^\star}$Present address: National Institute of Standards and Technology (NIST), 325 Broadway, Boulder, CO 80305, USA}

\maketitle
%\twocolumn
%\tableofcontents
\section{Introduction}

Since 1967 the time unit "second" of the international system of units (SI) is defined using the ground state hyperfine transition of the caesium atom $^{133}$Cs. Today a set of highly accurate atomic caesium fountains, some of them reaching frequency uncertainties in the low 10$^{-16}$, realizes the present definition of the SI-second with the lowest uncertainty. Besides the utilization for local time scale steering \cite{bauch2012, rovera2016}, fundamental research (e.g. \cite{guena2012b,huntemann2014,hees2016}) and measurements of optical frequencies (e.g. \cite{grebing2016, lodewyck2016}), the main task of primary fountain clocks is the calibration of the widely used International Atomic Time (TAI) calculated by the Bureau International des Poids et Mesures (BIPM). 
Our two institutes, Laboratoire National de m\'{e}trologie et d'Essais - SYst\`{e}me de R\'{e}f\'{e}rences Temps-Espace (LNE-SYRTE) in Paris as Designated Institute for France and Physikalisch-Technische Bundesantsalt (PTB) in Braunschweig as National Metrology Institute (NMI) for Germany, maintain several atomic fountains which are among the world's best primary clocks regarding both accuracy and reliability, i.e. capability of almost continuous operation.   They are used to steer timescales at LNE-SYRTE and PTB on a daily basis, which yields some of the most stable local real time realizations of the Coordinated Universal Time (UTC) \cite{bauch2012, rovera2016}.
Over the last decade, these fountains have provided to the BIPM a large number of formal monthly calibration reports with stated relative uncertainties of a few 10$^{-16}$, representing about 60-70\% of the total number of calibrations published in the \textit{Circular T}~\cite{circulart}.   
Given the major involvement of these fountains in the accuracy of TAI, it is crucial to test their stated uncertainties. So far testing the uncertainty of a clock was mainly based on local clock comparisons, which was enabled both at LNE-SYRTE and at PTB with three and two simultaneously operating caesium atomic fountains, respectively. A more stringent test is to compare distant clocks operated by independent teams in separate laboratories. To date, however, the uncertainties of distant comparisons have been limited by the comparison methods relying on satellite based time and frequency transfer techniques via the U.S. Global Positioning System (GPS) or Two-Way Satellite Time and Frequency Transfer (TWSTFT) \cite{bauch2005, fujieda2016}, or TAI itself used as a flywheel \cite{parker2010,parker2012, petit2013}.
 
In this paper, we report the first comparison of distant fountains, six pairs with three fountains at LNE-SYRTE and two at PTB, at the limit of the fountain uncertainties using a 1415\,km long stabilized optical fiber link between Paris and Braunschweig. This link was specially developed for high resolution frequency comparisons of optical clocks from our two institutes \cite{chiodo2015,raupach2015,lisdat2016} and has been characterized to contribute a frequency instability of $\sim$10$^{-15}$ at 1\,s averaging time. The fountain comparison, carried out in June 2015, demonstrates that  frequency transfer by means of an optical fiber link is also highly beneficial to remote comparisons of fountain clocks.

\section{Setup}\label{sec_setup}

\begin{figure}[h]
\includegraphics [width= 1.0\linewidth]{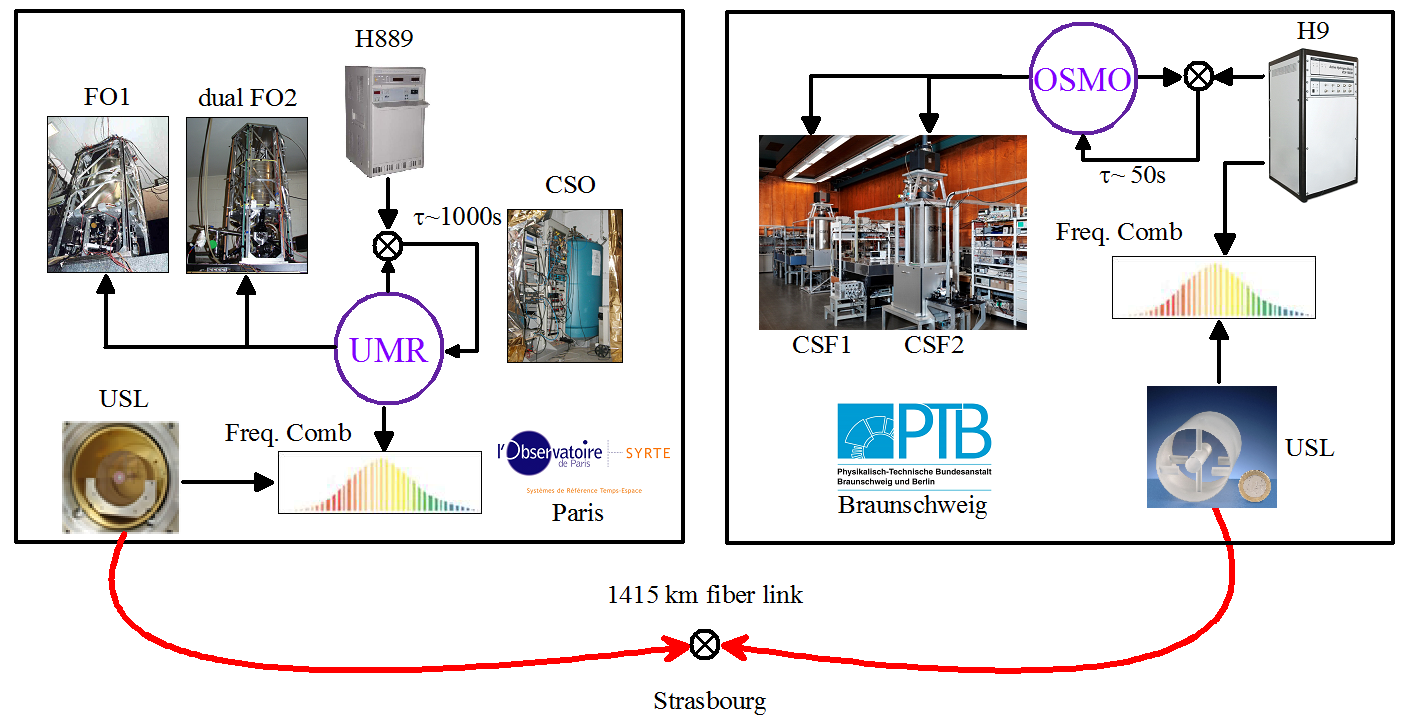}
\caption{Schematic view of the experiment. UMR: Ultrastable Microwave Reference; CSO: Cryogenic Sapphire Oscillator; OSMO: Optically Stabilized Microwave Oscillator; USL: Ultrastable Laser. FO1, CSF1 and CSF2 are caesium fountains while FO2 is a dual fountain operating with caesium and rubidium simultaneously. H889 and H9 are active hydrogen masers.}
\label{fig_setup}
\end{figure}

A general overview of the setup is given in Fig.~\ref{fig_setup}. At each institute, a set of atomic fountains measures the frequency of a hydrogen maser, filtered by a low phase noise microwave oscillator (an Ultrastable Microwave Reference (UMR) at SYRTE \cite{guena2012} and an Optically Stabilized Microwave Oscillator (OSMO) at PTB \cite{Lipphardt2016}). The hydrogen maser frequencies are linked to ultrastable frequency transfer lasers (USLs) at $1.5\,\mu$m using optical frequency combs (OFCs). The USLs are injected into stabilized telecommunication fibre links to the connection point in Strasbourg \cite{chiodo2015,raupach2015}.

\subsection{Fountains and local oscillators}\label{subsec_fountains}

The fountains  intercompared are the primary frequency standards (PFS) FO1, FO2 located at SYRTE \cite{guena2012} and CSF1, CSF2 \cite{weyers2001,gerginov2010} located at PTB. FO2 is a dual fountain operating simultaneously with caesium and rubidium atoms, known as FO2-Cs (PFS) and FO2-Rb (secondary frequency standard realizing the $^{87}$Rb secondary representation of the second). FO2-Rb, which has comparable performance as the PFSs, also regularly contributes to the steering of TAI \cite{guena2014}. At SYRTE the local interrogation signal is derived from the UMR which is based on a cryogenic sapphire oscillator and phase-locked to the hydrogen maser H889 with a time constant of $\sim\,1000\,$s \cite{guena2012}. At PTB the local interrogation signal is obtained from the OSMO, locked to the maser H9 with a time constant of $\sim\,50\,$s \cite{Lipphardt2016}.      
  
The strategies to reduce and evaluate the frequency shifts and the corresponding uncertainties can be found in \cite{guena2012, guena2014, guena2011} and \cite{weyers2001, gerginov2010, weyers2001a, weyers2012} for SYRTE's and PTB's fountains, respectively. In this work, the relevant fountain accuracy budgets are close to those reported to the BIPM for the June 2015 evaluations (\textit{Circular~T330} \cite{circulart}).
The only exception is the fountain PTB-CSF1, for which a reevaluation of the distributed cavity phase shift has been performed after the comparison campaign, which retrospectively leads to a significantly reduced overall systematic uncertainty \cite{weyers2017}. 

In addition, for the distant fountain comparison we apply improved relativistic gravitational redshift corrections with reduced uncertainties. Recently in the framework of the International Timescales with Optical Clocks (ITOC, part of the European Metrological Research Programme EMRP) project, the gravity potentials at local reference markers at SYRTE and PTB were newly determined with respect to a common reference potential. This involved a combination of GPS based height measurements, geometric levelling and a geoid model \cite{denker2013}, refined by local gravity measurements. Besides, in both laboratories, the gravity potential differences between the local reference markers and the reference points of each fountain (i.e. the mean effective position of the moving atoms during their ballistic flight above the microwave cavity centre) were determined by geometric levelling and from the respective fountain geometries and launch velocities.
The uncertainty of the differential redshift determinations for the distant fountains is less than $4\times 10^{-18}$, corresponding to less than $4\,$cm uncertainty in height difference, which is insignificant compared to the overall systematic (type B) uncertainty of each fountain as listed in Table~\ref{tab_summary}.

\subsection{Transfer oscillators and optical link}\label{subsec_oscillators}

At each institute an optical frequency comb is used to measure the microwave reference (the UMR at SYRTE and H9 at PTB) frequency with respect to the frequency of an ultrastable infrared laser (USL) at $1.542\,\mu$m, serving as transfer laser between the two institutes. 
The combs are generated by mode-locked erbium-doped fiber lasers delivering femtosecond pulses centered around $1.5\,\mu$m. Mode-locked fiber lasers are demonstrated to support measurements lasting several weeks, given the robustness of the mode-lock and the very low degree of necessary maintenance.% $5\times10^{-16}$ in the $0.1-100\,$s range) into the repetition rate $f_{\mathrm{rep}}$ of the comb \cite{lodewyck2016}.
At SYRTE, the comb is locked to the transfer laser via a fast feedback loop, effectively transferring the spectral purity of the transfer laser (instability of $5\times10^{-16}$ in the $0.1-100\,$s range) into the repetition rate $f_{\mathrm{rep}}$ of the comb \cite{lodewyck2016}. 
To obtain the frequency ratio between the transfer laser and the UMR, the latter is transmitted to the comb via a compensated fiber link with an instability of a few $10^{-15}$ at $1\,$s. The frequency difference between the 36$^{\mathrm{th}}$ harmonic of the comb and the UMR is counted by a zero dead-time counter referenced to the UMR. At PTB, the carrier-envelope offset frequency $f_{\mathrm{CEO}}$ and the repetition rate $f_{\mathrm{rep}}$ of the comb are phase locked with low bandwidth to the hydrogen maser H9. For the actual measurement of the frequency ratio between the USL and H9, the transfer oscillator concept~\cite{telle2002} is employed. 

To combine the locally measured optical to microwave frequency ratios, we took advantage of the fiber link designed for the comparison of optical clocks between our two institutes and described in detail in \cite{chiodo2015,raupach2015,lisdat2016}. At each institute the transfer laser is injected into a bi-directional long-haul coherent fibre link with active compensation of propagation phase noise and partial loss compensation. At the University of Strasbourg, each of the two links feeds a repeater laser station (RLS) \cite{chiodo2015} regenerating the optical signals. The optical beat notes between the user's outputs of the two RLSs are recorded with a dead-time free frequency counter referenced to an ultra-stable quartz disciplined by GPS. 

Given the simultaneous frequency measurements in Paris, Strasbourg and Braunschweig, the frequency difference between the remote microwave references UMR$-$H9 can be calculated with negligible uncertainty contributions from the long-distance fiber link, the combs and the transfer lasers. The $1\,$s gate intervals of the frequency counters were synchronized to local representations of timescales UTC(k) or to GPS time. Additionaly, we actively 
cancelled the mid- and long-term drifts of the transfer lasers. Thus the uncertainty contribution due to synchronization issues is below $10^{-19}$ (details see \cite{lisdat2016}).

\section{Data analysis and results} \label{sec_results}

The measurement campaign was performed in June 2015, between MJDs 57177 and 57198 (MJD: Modified Julian Date). The evaluation of the link measurements produces data of the frequency difference between the UMR phase-locked to the maser H889 at SYRTE and the maser H9 at PTB, sampled at $1\,$s. At SYRTE the FO1, FO2-Cs and FO2-Rb fountain data (UMR$-$FOx fractional frequency differences) are available on the clock cycle basis of $\sim\,1.4\,$s$-1.6\,$s. The data processing of the PTB's fountains CSF1 and CSF2 provides fractional frequency differences (H9$-$CSFx) averaged over 1~hour intervals. The fountain measurements covered $>80\%$ of the campaign duration at SYRTE and $>90\%$ at PTB. The masers distant comparison via the link was in operation during about $70\%$ of the $\sim\,21~$days period.

The comparisons were performed over synchronous time intervals in order to cope with gaps in the data, for both the four local and the six remote fountain pairs available. For the local comparisons at SYRTE, the fountain data were averaged over $100\,$s intervals, much longer than the clock cycle times. The data of the PTB's fountains provided over synchronous 1~hour long intervals could be directly intercompared. In view of the SYRTE to PTB comparisons, SYRTE's fountain data and the link data were also averaged over 1~hour. Only 1 hour intervals containing more than 25\% valid data from the link as well as the participating fountains were included.
The results were cross-checked using various evaluation routines independently developed at SYRTE and PTB.

\begin{figure}[h]
\hspace{-10mm}
\includegraphics [width= 0.6\linewidth]{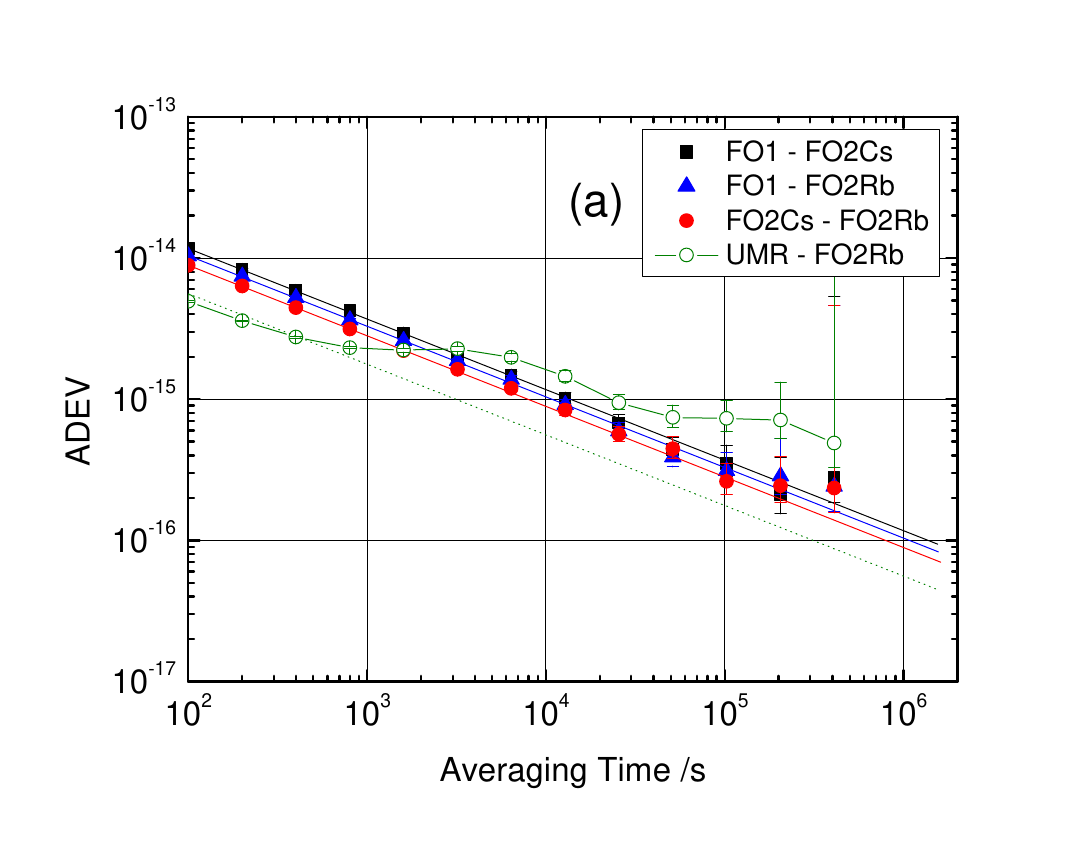}
\hspace{-10mm}
\includegraphics [width= 0.6\linewidth]{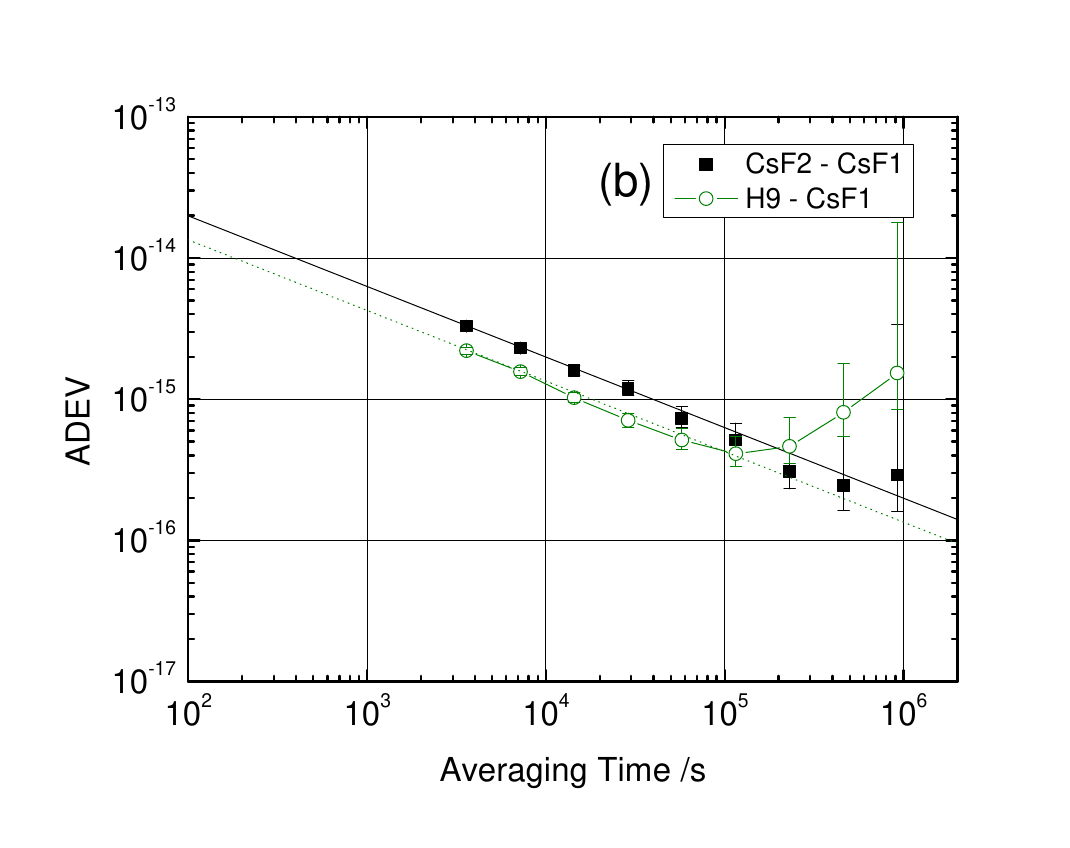}
\caption{Allan standard deviations (ADEV) of local comparisons (MJD 57177$-$57198) at SYRTE (a) and PTB (b). The straight lines correspond to extrapolations of the ADEV assuming white frequency noise. The open circles correspond to the frequency comparison between the local reference and one fountain.}
\label{LocalComp}
\end{figure}

Figures~\ref{LocalComp}(a) and \ref{LocalComp}(b) show the Allan standard deviation (ADEV) of the local fountain comparisons at SYRTE and PTB, respectively. The frequency instability for an averaging time $\tau$ is about $1\times 10^{-13} \tau ^{-1/2}$ for the three SYRTE's fountain pairs and about $2\times 10^{-13} \tau ^{-1/2}$ for CSF1$-$CSF2. Since all fountains are operated in the quantum projection noise limited regime [28], the individual instabilities are given by the respective detected atom numbers. The four local comparisons reach $2\times 10^{-16}$ after averaging periods of about 3 days at SYRTE and 10 days at PTB. The green curves in Figures~\ref{LocalComp}(a) and \ref{LocalComp}(b) give the ADEV of the local comparison between the frequency reference and one fountain at SYRTE and PTB, respectively. The UMR$-$FO2-Rb instability of $5\times 10^{-14} \tau ^{-1/2}$ for $\tau$ less than $1000\,$s is quantum projection noise limited thanks to the use of the cryogenic sapphire oscillator \cite{santarelli1999}. It then increases due to the phase locking of the UMR to the H889 maser and converges to the instability of H889. The H9$-$CSF1 short term instability of $1.2\times 10^{-13} \tau ^{-1/2}$ 
results from the CSF1 quantum projection noise and the H9 frequency instability. Beyond two days, the hydrogen maser frequency drifts ($\sim\,-1.5\times 10^{-16}\,$d$^{-1}$ and $\sim\,-1.9\times 10^{-16}\,$d$^{-1}$ for H889 and  H9, respectively) show up, but these are well rejected in the fountain synchronous differences, as expected for common noise.

\begin{figure}[h]
\hspace{-10mm}
\includegraphics [width= 0.6\linewidth]{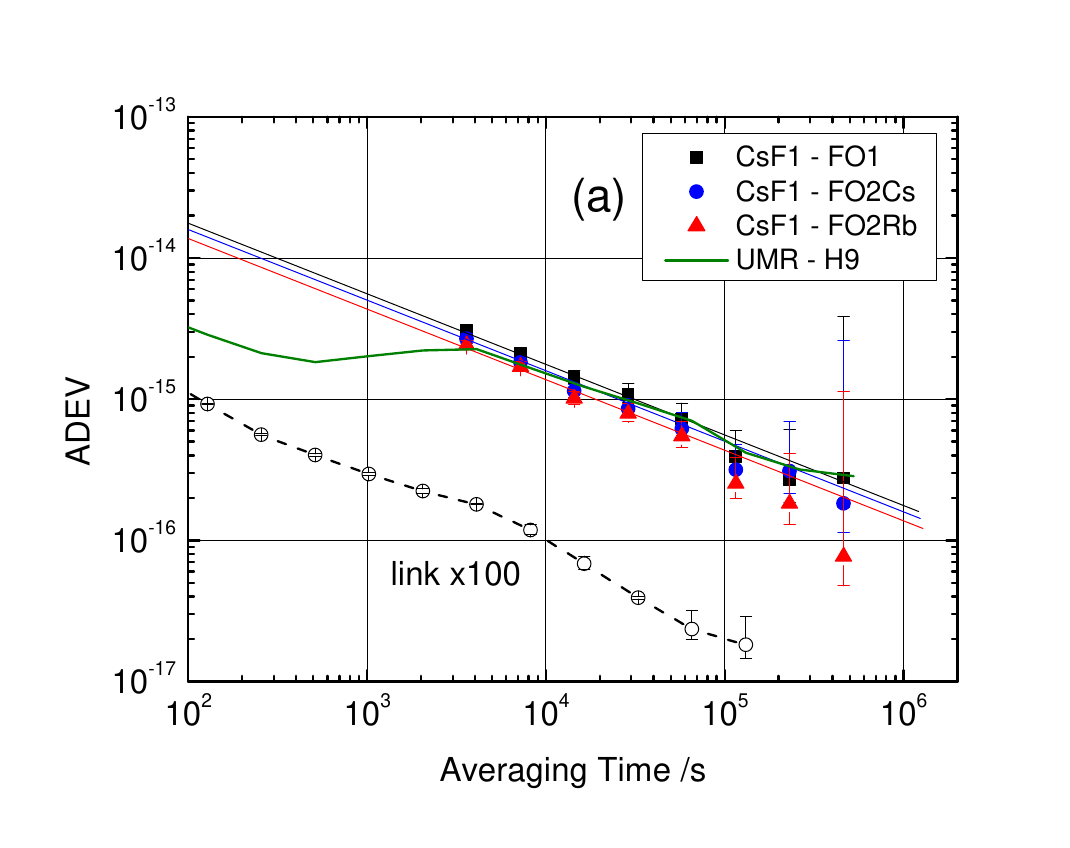}
\hspace{-10mm}
\includegraphics [width= 0.6\linewidth]{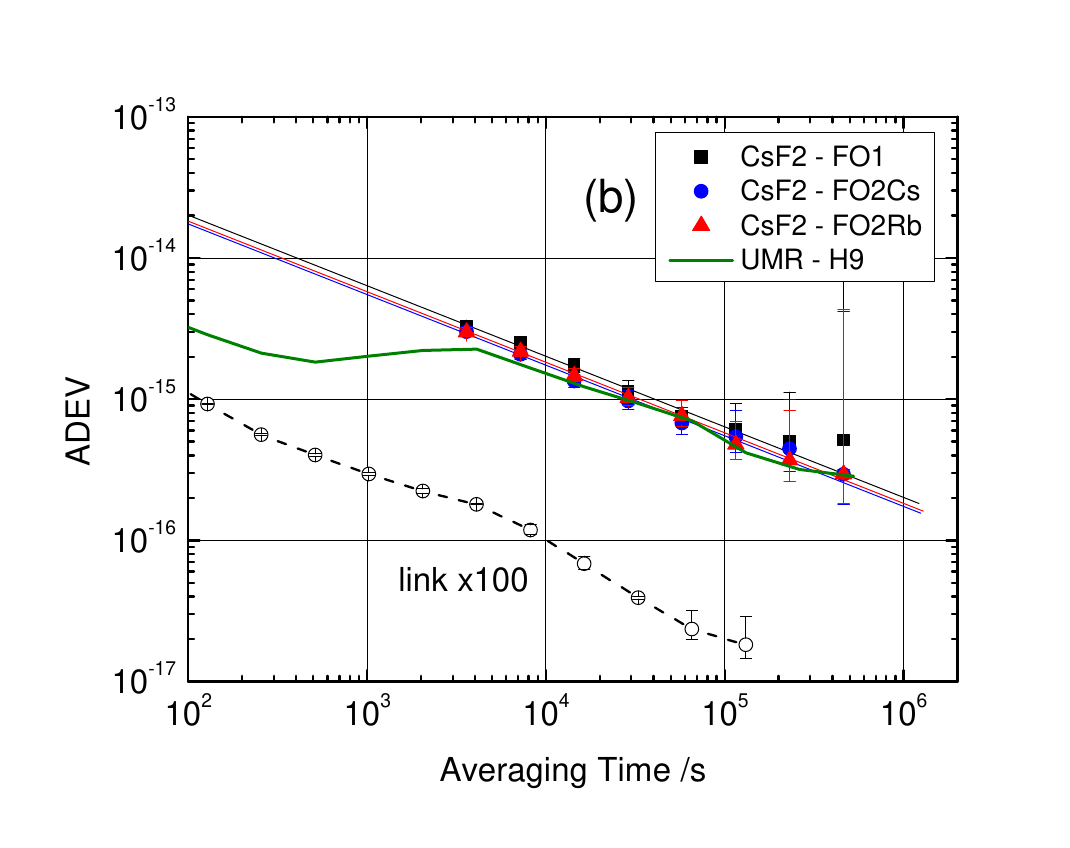}
\caption{Allan standard deviations (ADEV) of the distant fountain comparisons (MJD 57177$-$57198) (a) PTB-CSF1$-$SYRTE-FOx and (b) PTB-CSF2$-$SYRTE-FOx. The straight lines correspond to extrapolations of the ADEV assuming white frequency noise. The instability of the remote comparison between the local frequency references, the UMR (ultrastable microwave reference locked to H889 maser) at SYRTE and H9 maser at PTB, is shown by the dark green line and an upper limit of the instability of the fiber link \cite{lisdat2016}, multiplied by 100, by the open circle dashed line.}

\label{DistantComp}
\end{figure}

Figures~\ref{DistantComp}(a) and \ref{DistantComp}(b) present the ADEV of the distant comparisons of the SYRTE fountains against PTB$-$CSF1 and PTB$-$CSF2, respectively. The instability for all the comparisons is at or below $2\times 10^{-13}\,\tau\,^{-1/2}$ and reaches $2\times\,10^{-16}$ for averaging periods of $10\,$d or even less, just as if the fountains were located in the same laboratory (Fig.~\ref{LocalComp}). This result, remarkable considering the $1415\,$km signal path between the fountains, is obtained thanks to the very low noise of both the fiber link (open circle dashed line in Fig.~\ref{DistantComp}) and the comparison chains. Also shown in Figure~\ref{DistantComp} (green curve) is the instability of the distant comparison between the local frequency references UMR$-$H9. Below a few hundred seconds, it is limited by the noise of H9, the noise of the UMR at SYRTE being much smaller ($\sim\,2\times\,10^{-15}$ at $1\,$s). 
The instability decreases up to a few hundred seconds, as expected from the time constant of the UMR phase lock loop to H889 of about $1000\,$s. Then the ADEV follows that of the comparison of the H-masers which decreases up to averaging times of $10\,$d, because the drifts of the two masers happen to be very similar. 

\begin{table}
%\renewcommand{\arraystretch}{0.9}
\begin{center}
\begin{tabular}{lcccccc}
\hline%\br %Metrologia sty
(a) Local Comp.	&Diff. [$10^{-16}$]&	$u_{A}$[$10^{-16}$]	& $u_{B1}$[$10^{-16}$]	&$u_{B2}$[$10^{-16}$]	&  $u$[$10^{-16}$] & up-time [$\%$]\\
\hline
CSF2$-$CSF1	&-2.4	& 1.4	& 3.0	& 3.0   & 4.5 & 93\\
FO1$-$FO2-Cs	&-1.8	& 0.9	& 3.6	& 2.5	& 4.5 & 81\\
FO1$-$FO2-Rb	&-2.9	& 1.2	& 3.6	& 2.7	& 4.7 & 82 \\
FO2Cs$-$FO2-Rb	&-0.2	& 0.7	& 2.5	& 2.7	& 3.7 & 84\\
\hline %\mr
(b) Distant Comp.	& Diff.	& $u_{A}$	& $u_{B1}$	& $u_{B2}$	& $u$\\
\hline %\mr
CSF1$-$FO1	    & 2.7	& 1.6	& 3.0	& 3.6	& 5.0 & 68 \\
CSF1$-$FO2-Cs	& 0.6	& 1.4	& 3.0	& 2.5	& 4.2 & 70\\
CSF1$-$FO2-Rb	& 1.2	& 1.2	& 3.0	& 2.7	& 4.2 &72\\
CSF2$-$FO1	    &-0.1	& 1.8	& 3.0	& 3.6	& 5.0 & 68\\
CSF2$-$FO2-Cs	&-1.8	& 1.6	& 3.0	& 2.5	& 4.2 & 70\\
CSF2$-$FO2-Rb	&-1.9	& 1.6	& 3.0	& 2.7	& 4.3 &72\\
\hline
\end{tabular}
\end{center}
\caption{Summary of the local and distant comparisons between the SYRTE and PTB fountains during the June 2015 campaign (MJD 57177-57198). The second column gives the average fountain difference (Diff.) $u_{A}$ is the statistical uncertainty corresponding to the Allan standard deviation presented in Figures \ref{LocalComp} and \ref{DistantComp} extrapolated at the measurement duration assuming white frequency noise. $u_{B1}$ and $u_{B2}$ correspond to the systematic uncertainty of each fountain under comparison. $u$ is the combined uncertainty. The last column lists the up-time of the comparisons. The value of the absolute frequency of the $^{87}$Rb ground state hyperfine transition used in these comparisons is 6~834~682~610.904~312~Hz. In 2015 this was the value recommended by the Comit\'{e} International des Poids et Mesures.} 
\label{tab_summary}
\end{table}

The measured mean fractional frequency differences for all the fountain pairs are listed in Table~\ref{tab_summary}, together with the estimated statistical uncertainties $u_{A}$, the systematic uncertainties $u_{B1}$ and $u_{B2}$ of the two fountains compared, and the up-times of the comparisons over the period 57177$-$57198 ($4^{\mathrm{th}}-25^{\mathrm{th}}$ June 2015). The effective comparison durations range from 14 to 15\,d over the 3 weeks campaign. The combined uncertainties $u$, quadratic sums of the three uncertainties, are dominated by the type B uncertainties and range from  $3.7\times 10^{-16}$ to  $5.0\times 10^{-16}$. The differences ranging between $-2.9\times 10^{-16}$ and  $+2.7\times 10^{-16}$ indicate agreement between the fountains within these uncertainties. Finally, we note that the agreement of the remote fountains as observed in table \ref{tab_summary} is further supported by recent absolute optical frequency measurements of the strontium optical clock transition performed independently at PTB and at SYRTE \cite{grebing2016,lodewyck2016}. The related results agree to better than $3\times 10^{-16}$ with an uncertainty of $3.8\times 10^{-16}$ dominated by the fountain clock uncertainty contributions. 

Additionally, the comparisons of the four Cs fountain PFSs to FO2-Rb provide new absolute determinations of the $^{87}$Rb ground state hyperfine transition frequency.  This transition is recognized as a secondary representation of the second (SRS) and the fountain FO2-Rb regularly contributes to the calibration of TAI. The four results (see Table 1) agree well within the combined statistical ($u_{A}$ in column 3) and systematic caesium fountain uncertainties ($u_{B1}$ in column 4).
Considering these combined uncertainties to be independent to a large extent, we use them to calculate the weighted mean of the comparisons and its uncertainty (the latter by quadratically adding the systematic uncertainty of the rubidium fountain). The two local comparisons at SYRTE (Table 1(a), rows 3 and 4) thus lead to 6~834~682~610.904~3127~(24)\,Hz  and the two distant comparisons (Table 1(b), rows 3 and 6) to 6~834~682~610.904~3122~(24)\,Hz. Combining the four comparisons accordingly, we get our best new determination 6~834~682~610.904~3125~(21)\,Hz ($3.1\times 10^{-16}$ relative uncertainty) of the $^{87}$Rb hyperfine transition frequency. Our results agree with the current definition 6~834~682~610.904~310~Hz of the $^{87}$Rb SRS that was adopted at the last meeting of the Comit\'{e} International des Poids et Mesures (CIPM) \cite{CIPM2015} with a recommended relative uncertainty of $7\times 10^{-16}$ (4.8 $\mu$\,Hz).

\section{Conclusion}\label{sec_conclusion}

Utilizing a phase coherent optical fiber link, we have performed a remote frequency comparison of primary and secondary frequency standards, caesium and rubidium fountains, located in metrology laboratories 700\,km apart. To our knowledge this is the first remote comparison of primary frequency standards via an optical fiber link. The comparison interval spanned about 21 days in June 2015. The measured frequency differences for the four individual pairs of distant Cs fountains are all in the range $\pm 3\times 10^{-16}$ which is fully compatible with the combined uncertainties. These results support the performance and stated accuracies of the SYRTE's and PTB's fountain PFSs, which is important for their utilization in fundamental physics and as primary frequency standards in metrology. We also provide a new absolute determination of the $^{87}$Rb ground state hyperfine transition frequency exploiting the new capabilities of frequency transfer afforded by an ultrastable optical fiber link.  
It is expected to repeat such comparisons, and extend them to some other laboratories within the European fiber network currently under development.   

%\footnote{${^\star}$Present address: National Institute of Standards and Technology (NIST), 325 Broadway, Boulder, CO 80305, USA}

%V.~Gerginov and D.~Nicolodi are now with National Institute of Standards and Techno-logy (NIST), 325 Broadway, Boulder, CO 80305, USA.\\

\section*{Acknowledgments}

We are grateful for the constant technical support, both by SYRTE's technical services, and at PTB by M.~Kazda and A.~Koczwara, thanking T.~Legero and U.~Sterr for operating the cavity-stabilised laser. We thank F.~Riehle and H.~Schnatz at PTB for their long-standing support. We would like to thank P.~Delva from SYRTE, and L.~Timmen and C.~Voigt from the Leibniz Universit\"{a}t Hannover for supporting the new evaluation of the gravity potential at the sites of SYRTE and PTB. We thank P.~Gris and B.~Moya for their help establishing the cross-border link between Kehl and Strasbourg, and C.~Grimm from Deutsches Forschungsnetz (DFN) and W.-Ch.~K{\"{o}}nig (Gasline GmbH) for facilitating network access.
We acknowledge funding support from the Agence Nationale de la Recherche (ANR blanc LIOM 2011-BS04-009-01, Labex First-TF ANR-10-LABX-48-01, Equipex REFIMEVE+ ANR-11-EQPX-0039), Centre National d'\'{E}tudes Spatiales (CNES), Conseil R\'{e}gional Ile-de-France (DIM Nano'K), CNRS with Action Sp\'{e}cifique GRAM, 
the Deutsche Forschungsgemeinschaft (DFG) within the CRC 1128 (geo-Q, projects A04, C03 and C04), the European Metrology Research Programme (EMRP) in the Joint Research Projects SIB02 (NEAT-FT) and SIB55 (ITOC), and the European Metrology Programme for Innovation and Research (EMPIR) in project 15SIB05 (OFTEN). The EMRP and EMPIR are jointly funded by the EMRP/EMPIR participating countries within EURAMET and the European Union.

\newpage

\section*{References} %metrologia sty


\begin{thebibliography}{16}

\providecommand{\url}[1]{#1}
\csname url@samestyle\endcsname

\bibitem{bauch2012}
Bauch A, Weyers S, Piester D, Staliuniene E and Yang W, ``Generation of UTC(PTB) as a fountain-clock based time scale,'' 2012 \emph{Metrologia} \textbf{49} 180-188

\bibitem{rovera2016}
Rovera G D, Bize S, Chupin B, Gu{\'e}na J, Laurent Ph, Rosenbusch P, Uhrich P and Abgrall M, ``UTC(OP) based on LNE-SYRTE atomic fountain primary frequency standards,''  2016 \emph{Metrologia} \textbf{53} S81-S88

\bibitem{guena2012b}
Gu{\'e}na J, Abgrall M, Rovera D, Rosenbusch P, Tobar M E, Laurent Ph, Clairon A and Bize S,''Improved Tests of Local Position Invariance Using $^{87}$Rb and $^{133}$Cs Fountains" 2012 \emph{Phys. Rev. Lett.} \textbf{109} 080801

\bibitem{huntemann2014}
Huntemann N, Lipphardt B, Tamm Chr, Gerginov V, Weyers S, and Peik E, ``Improved Limit on a Temporal Variation of m$_{p}$/m$_{e}$ from Comparisons of Yb$^{+}$ and Cs Atomic Clocks'' 2014 \emph{Phys. Rev. Lett.} \textbf{113} 210802

\bibitem{hees2016}
Hees A, Gu{\'e}na J, Abgrall M, Bize S, and Wolf P, ``Searching for an Oscillating Massive Scalar Field as a Dark Matter Candidate Using Atomic Hyperfine Frequency Comparisons''  2016 \emph{Phys. Rev. Lett.} \textbf{117} 061301

\bibitem{grebing2016}
Chr.~Grebing \emph{et al.}, ``Realization of a timescale with an accurate optical lattice clock'' 2016 \emph{Optica} \textbf{3} 563-569

\bibitem{lodewyck2016}
Lodewyck J \emph{et al.} ``Optical to microwave clock frequency ratios with a nearly continuous strontium optical
lattice clock'' 2016 \emph{Metrologia} \textbf{53} 1123-1130

\bibitem{circulart}
www.bipm.org/jsp/en/TimeFtp.jsp

\bibitem{bauch2005}
Bauch A \emph{et al.} ''Comparison between frequency standards in Europe and the USA at the $10^{-15}$ uncertainty level'' 2006 \emph{Metrologia} \textbf{43} 109-120

\bibitem{fujieda2016}
Fujieda M, Achkar J, Riedel F, Takiguchi H, Benkler E, Abgrall M, Gu{\'e}na J, Weyers S and Piester D,``Carrier-Phase Two-Way Satellite Frequency Transfer between LNE-SYRTE and PTB'' 2016 \emph{Proc. of 30th Eur. Freq. \& Time Forum} April 2016 (York, U.K.)

\bibitem{parker2010}
Parker T E ``Long-term comparison of caesium fountain primary frequency
  standards'' 2010 \emph{Metrologia} \textbf{47} 1 [Online]. Available:
  \url{http://stacks.iop.org/0026-1394/47/i=1/a=001}

\bibitem{parker2012}
Parker T E ``Invited review article: The uncertainty in the realization and
  dissemination of the {SI} second from a systems point of view'' 2012 \emph{Review
  of Scientific Instruments} \textbf{83} 021102

\bibitem{petit2013}
Petit G and Panfilo G ``Comparison of frequency standards used for TAI'' 2013
  \emph{IEEE Trans. Instrum. Meas.} \textbf{62} 1550-1555

\bibitem{chiodo2015}
Chiodo N \emph{et al.} ``Cascaded optical fiber link using the internet network for remote clocks comparison,'' 2015 \emph{Opt. Express} \textbf{23} 33927-33937 

\bibitem{raupach2015}
Raupach SMF, Koczwara A, Grosche G ``Brillouin amplification supports $1\ifmmode\times\else\texttimes\fi{}{10}^{\ensuremath{-}20}$ uncertainty in optical frequency transfer over 1400 km of underground fiber,'' 2015 \emph{Phys. Rev. A} \textbf{92} 021801(R)

\bibitem{lisdat2016}
Lisdat C \emph{et al.} ``A clock network for geodesy and fundamental science,'' 2016 \emph{Nat. Commun.} \textbf{7} 12443

\bibitem{guena2012}
Gu{\'e}na J \emph{et al.} ``Progress in atomic fountains at LNE-SYRTE'' 2012 \emph{IEEE Trans. Ultrason. Ferroelectr. Freq. Control} \textbf{59} 391-410

\bibitem{Lipphardt2016}
Lipphardt B, Gerginov V, Weyers S, ``Optical Stabilization of a Microwave Oscillator for Fountain Clock Interrogation'' arxiv.org/abs/1609.05718

\bibitem{weyers2001}
Weyers S, H{\"u}bner U, Schr{\"o}der R, Tamm Chr and Bauch A ``Uncertainty evaluation of the atomic caesium fountain CSF1 of PTB'' 2001 \emph{Metrologia} \textbf{38} 343-352

\bibitem{gerginov2010}
Gerginov V, Nemitz N, Weyers S, Schr\"{o}der R, Griebsch D and Wynands R
  ``Uncertainty evaluation of the caesium fountain clock PTB-CSF2'' 2010 \emph{Metrologia} \textbf{47} 65-79

\bibitem{guena2014}
Gu{\'e}na J, Abgrall M, Clairon A and Bize S ``Contributing to TAI with a secondary representation of the second'' 2014 \emph{Metrologia} \textbf{51} 108-120

\bibitem{guena2011}
Gu{\'e}na J, Li R, Gibble K, Bize S and Clairon A ``Evaluation of Doppler Shifts to improve the accuracy of atomic fountain clocks'' 2011 \emph{Phys. Rev. Lett.} \textbf{106} 130801

\bibitem{weyers2001a}
Weyers S, Bauch A, Schr\"{o}der R and Tamm Chr ``The atomic caesium fountain CSF1 of PTB'' 2001 \emph{Proceedings of the 6th Symposium on Frequency Standards and Metrology} University of St Andrews, Fife, Scotland, 64-71, ISBN 981-02-4911-X (World Scientific)

\bibitem{weyers2012}
Weyers S, Gerginov V, Nemitz N,  Li R and Gibble K  ``Distributed cavity phase frequency shifts of the caesium fountain PTB-CSF2'' 2012 \emph{Metrologia} \textbf{49} 82-87

\bibitem{weyers2017} to be published

\bibitem{denker2013}
Denker H in Science of Geodesy-II (ed. G. Xu), 185-291, Springer Verlag, Berlin, Heidelberg, 2013

\bibitem{telle2002}
Telle H, Lipphardt B and Stenger J, ''Kerr-lens, mode-locked lasers as transfer oscillators for optical frequency measurements'' 2002 \emph{Appl. Phys. B} \textbf{74} 1-6

\bibitem{santarelli1999}
Santarelli G, Laurent P, Lemonde P,  Clairon A, Mann A G, Chang S, Luiten A N and Salomon C  ''Quantum projection noise in an atomic fountain: A high stability cesium frequency standard'' 1999 \emph{Phys. Rev. Lett.} \textbf{82} 4619–4622

\bibitem{CIPM2015}
International Committee for Weights and Measures 2015, Recommendation 2 (CI-2015), Proceedings of Session II of the 104th meeting (15-16 October 2015), BIPM, p.~48 
  
\end{thebibliography}
\end{document}